\documentclass[sigconf]{acmart}
\copyrightyear{2017} 
\acmYear{2017} 
\setcopyright{acmcopyright}
\acmConference{ICTIR '17}{October 1--4, 2017}{Amsterdam, Netherlands}
\acmPrice{15.00}
\acmDOI{http://dx.doi.org/10.1145/3121050.3121072}
\acmISBN{ISBN 978-1-4503-4490-6/17/10}
\usepackage{booktabs} 
\usepackage{multirow}
\setcopyright{rightsretained}
\usepackage{amsmath}
\providecommand{\dotdiv}{
\mathbin{
    \vphantom{+}
    \text{
      \mathsurround=0pt 
      \ooalign{
        \noalign{\kern-.35ex}
        \hidewidth$\smash{\cdot}$\hidewidth\cr 
        \noalign{\kern.35ex}
        $-$\cr 
      }%
    }%
  }%
}

\begin{document}
\title{Evaluation Measures for Relevance and Credibility in Ranked Lists}
\author{Christina Lioma, Jakob Grue Simonsen}
\affiliation{Department of Computer Science\\
  \institution{University of Copenhagen}
  \city{Copenhagen, Denmark} 
}
\email{{c.lioma,simonsen}@di.ku.dk}
\author{Birger Larsen}
\affiliation{Department of Communication\\
  \institution{University of Aalborg in Copenhagen}
  \city{Copenhagen, Denmark} 
}
\email{birger@hum.aau.dk}
\begin{abstract}
Recent discussions on \textit{alternative facts}, \textit{fake news}, and \textit{post truth} politics have motivated research on creating technologies that allow people not only to \textit{access} information, but also to \textit{assess} the credibility of the information presented to them by information retrieval systems. Whereas technology is in place for filtering information according to relevance and/or credibility \cite{LiomaLLH16}, no single measure currently exists for evaluating the accuracy or precision (and more generally \textit{effectiveness}) of \textit{both} the relevance \textit{and} the credibility of retrieved results. One obvious way of doing so is to measure relevance and credibility effectiveness separately, and then consolidate the two measures into one. There at least two problems with such an approach: (I) it is not certain that the same criteria are applied to the evaluation of both relevance and credibility (and applying different criteria introduces bias to the evaluation); (II) many more and richer measures exist for assessing relevance effectiveness than for assessing credibility effectiveness (hence risking further bias).

Motivated by the above, we present two novel types of evaluation measures that are designed to measure the effectiveness of both relevance and credibility in ranked lists of retrieval results. Experimental evaluation on a small human-annotated dataset (that we make freely available to the research community) shows that our measures are expressive and intuitive in their interpretation. 
\end{abstract}

\keywords{relevance; credibility; evaluation measures}

\maketitle

\section{Introduction}
\label{s:intro}
Recent discussions on \textit{alternative facts}, \textit{fake news}, and \textit{post truth} politics have motivated research on creating technologies that allow people, not only to \textit{access} information, but also to \textit{assess} the credibility of the information presented to them \cite{EnnalsBAR10,LexKBG14}. In the broader area of information retrieval (IR), various methods for approximating \cite{HornZGKL13,WiebeR11} or visualising \cite{MorrisCRHS12,ParkKCS09,SchwarzM11,HuangOA13} information credibility have been presented, both stand-alone and in relation to relevance \cite{LiomaLLH16}. Collectively, these approaches can be seen as steps in the direction of building IR systems that retrieve information that is both relevant and credible. Given such a list of IR results, which are ranked decreasingly by both relevance and credibility, 
the question arises: how can we evaluate the quality of this ranked list?

One could measure  retrieval effectiveness first, using any suitable existing relevance measure, such as NDCG or AP, and then measure separately credibility accuracy similarly, e.g. using the F-1 or the G-measure. This approach would output scores by two separate metrics\footnote{In this paper, we use \textit{metric} and \textit{measure} interchangeably, as is common in the IR community, even though the terms are not synonymous. Strictly speaking, \textit{measure} should be used for more concrete or objective attributes, and \textit{metric} should be used for more abstract, higher-level, or somewhat subjective attributes \cite{BlackSS08}. When discussing effectiveness, which is generally hard to define objectively, but for which we have some consistent feel, Black et al. argue that the term \textit{metric} should be used \cite{BlackSS08}.}, which would need to somehow be consolidated or considered together when optimising system performance. In such a case, and depending on the choice of relevance and credibility measures, it would not be always certain that the same criteria are applied to the evaluation of both relevance and credibility. For instance, whereas the state of the art metrics in relevance evaluation treat relevance as graded and consider it in relation to the rank position of the retrieved documents (we discuss these in Section \ref{s:rw}), no metrics exist that consider graded credibility accuracy in relation to rank position. Hence, using two separate metrics for relevance and credibility may, in practice,  bias the overall evaluation process in favour of relevance, for which more thorough evaluation metrics exist.

To provide a more principled approach that obviates this bias, we present two new types of evaluation measures that are designed to measure the effectiveness of both relevance and credibility in ranked lists of retrieval results \textit{simultaneously} and \textit{without bias} in favour of either relevance or credibility. Our measures take as input a ranked list of documents, and assume that assessments (or their approximations) exist both for the relevance and for the credibility of each document. Given this information, our Type I measures define different ways of measuring the effectiveness of both relevance and credibility based on differences in the \textit{rank position} of the retrieved documents with respect to their ideal rank position (when ranked only by relevance or credibility). Unlike Type I, our Type II measures operate directly on \textit{document scores} of relevance and credibility, instead of rank positions. We evaluate our measures both axiomatically (in terms of their properties) and empirically on a small human-annotated dataset that we build specifically for the purposes of this work. We find that our measures are expressive and intuitive in their interpretation.

\section{Related Work}
\label{s:rw}
The aim of evaluation is to measure how well some method achieves its intended purpose. This allows to discover weaknesses in the given method, potentially leading to the development of improved approaches and generally more informed deployment decisions. 
For this reason, evaluation has been a strong driving force in IR, where, for instance, the literature of IR evaluation measures is rich and voluminous, spanning several decades. Generally speaking, relevance metrics for IR can be split into three high-level categories: 
\begin{itemize}
\item[(i)] earlier metrics, assuming binary relevance assessments; 
\item[(ii)] later metrics, considering graded relevance assessments, and 
\item[(iii)] more recent metrics, approximating relevance assessments from user clicks. 
\end{itemize}

We overview some among the main developments in each of these categories next.

\subsection{Binary relevance measures}
Binary relevance metrics are numerous and widely used. Examples include: 
\begin{description}
\item[Precision @ $k$ (P@$k$):] the proportion of retrieved documents that are relevant, up to and including position $k$ in the ranking;
\item[Average Precision (AP):] the average of (un-interpolated) precision values (proportion of retrieved documents that are relevant) at all ranks where relevant documents are found;
\item[Binary Preference (bPref):] this is identical to AP except that bPref ignores non-assessed documents (whereas AP treats non-assessed documents as non-relevant). Because of this, bPref does not violate the \textit{completeness assumption}, according to which ``all relevant documents within a test collection have been identified and are present in the collection'') \cite{BuckleyV04};
\item[Mean Reciprocal Rank (MRR):] the reciprocal of the position in the ranking of the first relevant document only;
\item[Recall:] the proportion of relevant documents that are retrieved;
\item[F-score:] the equally weighted harmonic mean of precision and recall.
\end{description}

\subsection{Graded relevance measures}
There exist noticeably fewer graded relevance metrics than binary ones. The two main graded relevance metrics are NDCG and ERR: 
\begin{description}
\item[Normalised Discounted Cumulative Gain (NDCG):] the \\cumulative gain a user obtains by examining the retrieval result up to a rank position, where the relevance scores of the retrieved documents are:
\begin{itemize}
\item \textit{accumulated} over all the rank positions that are considered, 
\item \textit{discounted} in order to devaluate late-retrieved documents, and 
\item \textit{normalised} in relation to the maximum score that this metric can possibly yield on an ideal reranking of the same documents. 
\end{itemize}
Two useful properties of NDCG are that it rewards retrieved documents according to both (i) their degree (or grade) of relevance, and (ii) their rank position. Put simply, this means that the more relevant a document is \textit{and} the closer to the top it is ranked, the higher the NDCG score will be \cite{JarvelinK02}.

\item[Expected Reciprocal Rank (ERR):] ERR operates on the \\same high-level idea as NDCG but differs from it in that it penalises documents that are shown below very relevant documents. That is, whereas NDCG makes the \textit{independence assumption} that ``a document in a given position has always the same gain and discount independently of the documents shown above it'', ERR does not make this assumption, and, instead, considers (implicitly) the immediate context of each document in the ranking. In addition, instead of the discounting of NDCG, ERR approximates the expected reciprocal length of time that a user will take to find a relevant document. Thus, ERR can be seen as an extension of (the binary) MRR for graded relevance assessments \cite{ChapelleMZG09}.
\end{description}

\subsection{User click measures}
The most recent type of evaluation measures are designed to operate, not on traditionally-constructed relevance assessments (defined by human assessors), but on approximations of relevance assessments from user clicks (actual or simulated). Most of these metrics have underlying user models, which capture how users interact with retrieval results. In this case, the quality of the evaluation measure is a direct function of the quality of its underlying user model \cite{YilmazSCR10}. 

The main advances in this area include the following:
\begin{description}
\item[Expected Browsing Utility (EBU):] an evaluation measure whose underlying user click model has been tuned by observations over many thousands of real search sessions \cite{YilmazSCR10};
\item[Converting click models to evaluation measures:] a general method for converting any click model into an evaluation metric \cite{ChuklinSR13}; and
\item[Online evaluation:] various different algorithms for interleaving \cite{SchuthHR15} or multileaving \cite{BrostCSL16a,BrostSCL16,SchuthOWR16} multiple initial ranked lists into a single combined ranking, and by approximating clicks (through user click models) on the resulting combined ranking, assigning credit (hence evaluating) the methods that produced each initial ranked list \cite{HofmannLR16}.
\end{description}

In addition to the above three types of IR evaluation measures, there also exists further literature on IR measures that consider additional dimensions on top of relevance, such as query difficulty for instance \cite{Mizzaro08}. To the best of our knowledge, none of these measures consider credibility. The closest to a credibility measure we could find is the work by Balakrishnan et al.\ \cite{BalakrishnanK11} on source selection for deep web databases: their method considers the agreement between different sources in answering a query as an indication of the credibility of the sources. An adjusted version of this agreement is modeled as a graph with vertices representing sources. Given such a graph, the credibility (or quality) of each source is calculated as the stationary visit probability of a random walk on this graph. 

The evaluation measures we present in Sections \ref{s:ranks} - \ref{s:scores} are the only ones, to our knowledge, that are designed to operate both on relevance and credibility. Beyond these two particular dimensions, reasoning more generally about different dimensions of effectiveness, the F-score, and its predecessor, van Rijsbergen's E-score \cite{keith74}, are early examples of a single evaluation measure combining two different aspects, namely precision and recall. We return to this discussion in Section \ref{s:scores}, where we present a variant of the F-score for aggregating relevance and credibility.

\section{Evaluation Desiderata}
\label{s:desiderata}
Given a ranked list of documents, the aim is to produce a measure that reflects how effective this ranking is with respect to \textit{both} the relevance of these documents to some query \textit{and} also the credibility of these documents (irrespective of a query).  

There are at least two basic ways to produce such a metric: \\
Either
\begin{itemize}
\item[(I)] gauge the difference in rank position(s) between an input ranking and ``ideal'' relevance and credibility rankings,
\end{itemize}
or
\begin{itemize}
\item[(II)] employ relevance and credibility \emph{scores} to gauge how well the input ranking reflects high versus low scores. 
\end{itemize}
Note that while (II) is reminiscent of existing measures for relevance
ranking, the fact that two distinct kinds of scores (relevance and credibility) -- perhaps having different ranges and behaviour -- must be combined may lead to further complications.

Accordingly, in the remainder of the paper, we call measures \emph{Type I} if they are based primarily on differences in rank position, and \emph{Type II} if they are based primarily on relevance and credibility scores.

Regardless of whether it is Type I or Type II, we reason that any measure must be easily interpretable. Hence, its scores should be normalised between $0$ and $1$, where low scores should indicate poor rankings, and high scores should indicate good rankings. The extreme points ($0$ and $1$) of the scale should preferably be attainable by particularly bad or particularly good rankings; as a minimum, if the ranking can be measured against an ``ideal'' ranking (as in, e.g. NDCG), the value $1$ should be attainable by the ideal ranking.

In addition to the above, there also exist desiderata for evaluation measures that are more debatable (e.g., how the measure should act in case of identical ranking scores for distinct documents). Below,
we list what we believe to be the most pertinent desiderata. The list encompasses desiderata tailored to evaluate measures that gauge ranking based on either \emph{rank position}
or on (relevance or credibility) \emph{scores}. For the desiderata pertaining to rank position, we need the following ancillary definition:

Let $D_i$ be a document at rank $i$. We then define an \textit{error} as any instance where  
\begin{itemize}
\item either (a) the relevance of a document at rank $i$ is greater than the relevance of a document at rank $i-1$, 
\item or (b) the credibility of a document at rank $i$ is greater than the credibility of a document at rank $i-1$.
\end{itemize}
This assumes that documents are ranked decreasingly by relevance and credibility, i.e. that the ``best'' document occurs at the lowest (i.e. first) rank.

We define the following eight desiderata (referred to as \textbf{D1}-\textbf{D8} henceforth):

\begin{description}
\item[D1] Larger errors should be penalised more than smaller errors;
\item[D2] Errors high in the ranking should be penalised more than errors low in ranking;
\item[D3] Let $\delta^r$ be the difference in relevance score between $D_i$ and $D_{i-1}$ when $D_{i-1}$ is more relevant than $D_{i}$. Similarly, let $\delta^c$ be the difference in credibility score between $D_i$ and $D_{i-1}$ when $D_{i-1}$ is more credible than $D_{i}$. Then, larger $\delta^r$ and $\delta^c$ values should imply larger error;
\item[D4] \textit{Ceteris paribus}, a credibility error on documents of high relevance should be penalised more than a credibility error on documents of low relevance;
\item[D5] The metric should be well-defined even if all documents have identical ranking/credibility scores;
\item[D6] Scaling the document scores used to produce the ranking by some constant should not affect the metric;
\item[D7] If all documents have the same relevance score, the metric should function as a credibility metric; and vice versa;
\item[D8] We should be able to adjust (by some easily interpretable parameter) how much we wish to penalise low credibility with respect to low relevance, if at all.
\end{description}

Next, we present two types of evaluation measures of relevance and credibility that satisfy (wholly or partially) the above desiderata: Type I measures (Section \ref{s:ranks}) operate solely on the \textit{rank positions} of documents; Type II measures (Section \ref{s:scores}) operate solely on \textit{document scores}.

\section{Type I: Rank Position Measures}
\label{s:ranks}
Given a ranking of documents that we want to evaluate (let us call this \textit{input ranking}), we reason in terms of two additional \textit{ideal rankings}: one by relevance only, and one by credibility only (the two ideal rankings are entirely independent of each other). So, for each document, we have: 
\begin{itemize}
\item[(1)] its rank position in the input ranking; 
\item[(2)] its rank position in the ideal relevance ranking; and 
\item[(3)] its rank position in the ideal credibility ranking. 
\end{itemize}
The basic idea is then to take each adjacent pair of documents in the input ranking, check for errors in the input ranking compared to the ideal relevance and separately the ideal credibility ranking, and aggregate those errors. We explain next how we do this.

Let $D_i$ be the rank position of document $D$ in the input ranking. We then denote by $R^r_ {D_i}$ the rank position of $D_i$ in the ideal relevance ranking, and by $R^c_{D_i}$ the rank position of $D_i$ in the ideal credibility ranking. 
Note that subscript $_i$ refers to the rank position of $D$ in the input ranking \textit{at all times}. That is, $R^r_ {D_i}$ should be read as: the position in the ideal relevance ranking of the document that is at position $i$ in the input ranking; similarly for $R^c_ {D_i}$. 
 
Let the \emph{monus operator} $\dotdiv$ be defined on non-negative real numbers by:
\begin{align}
a \dotdiv b = \left\{ \begin{array}{ll} 0 & \mathrm{if } \; a \leq b \\
					      a-b & \mathrm{if } \; a > b \end{array} \right.
\end{align}

\noindent That is, $a \dotdiv b$ is simply subtraction as long as $a > b$ and otherwise just returns $0$.
Then, using the monus operator and the notation introduced above, we define a ``relevance error'' ($\epsilon^r$) and a ``credibility error'' ($\epsilon^c$) as: 
\begin{equation}
\label{eq:error-r}
\epsilon^r = R^r_{D_{i}} \dotdiv R^r_{D_{i+1}}
\end{equation}
\begin{equation}
\label{eq:error-c}
\epsilon^c = R^c_{D_{i}} \dotdiv R^c_{D_{i+1}}
\end{equation}
In the above, $i$ and $i+1$ are the rank positions of two documents in the input ranking. 
Given two such documents, a ``relevance error'' occurs iff the document that is ranked lower (at rank $i$) in the input ranking is ranked after the other document in the ideal relevance ranking. Otherwise, the error is zero.
Similarly for the ``credibility error''.

For example, if three documents $A$, $B$ and $C$ are ranked as $C,A,B$ in the input ranking (i.e., $D_1 = C$, $D_2 = A$, $D_3 = B$), but ranked as $R^r = [A,B,C]$ in an ideal relevance ranking, there are two relevance errors, namely 
\begin{itemize}
\item[(i)] $R^r _{D_1} \dotdiv R^r_{D_2} = R^r_{C} \dotdiv R^r_{A} = 3 \dotdiv 1 = 2$, and
\item[(ii)] $R^r_{D_2} \dotdiv R^r_{D_3} = R^r_{A} \dotdiv R^r_{B} = 1 \dotdiv 2 = 0$.
\end{itemize}
We use the above ``relevance error'' and ``credibility error'' to define the two evaluation measures, presented next.

\subsection{Normalised Local Rank Error (NLRE)}
Let $n$ be the total number of documents in the ranked list.
We define the Local Rank Error (LRE) evaluation measure as $LRE = 0$ if $n = 1$, and otherwise:

\begin{equation}
\label{eq:lre}
LRE =   \sum_{i=1}^{n-1}  \frac{1}{\log_2 (1+i)} \left(\left(\mu + \epsilon^r \right) \left(\nu + \epsilon^c \right)
-\mu \nu \right)
\end{equation}

\noindent where $\epsilon^r, \epsilon^c$ are the relevance error and credibility error defined in Equations \ref{eq:error-r} -- \ref{eq:error-c}, and $\mu, \nu$ are non-negative real numbers (with $\mu + \nu > 0$) controlling how much we wish to penalise low relevance with respect to low credibility. For instance, a high $\nu$ weighs credibility more, whereas
a high $\mu$ weighs relevance more. The reason for the term $- \mu\nu$ inside the summation at the end is to ensure that the value of the LRE measure is zero if no error occurs.



Because Equation \ref{eq:lre} is \emph{large} for \emph{bad} rankings and \emph{small} for \emph{good} rankings, we invert and normalise it (Normalised LRE or NLRE) as follows:
\begin{equation}
\label{eq:nlre}
NLRE = 1 - \frac{LRE}{C_{LRE}}
\end{equation}

\noindent where $C_{LRE}$ is the normalisation constant, defined as:
\begin{equation} \label{eq:clre}
C_{LRE} = \sum_{j=0}^{\lfloor \frac{n}{2} - 1\rfloor} \frac{(n-2j-1)^2 + (\mu + \nu) (n - 2j - 1)}{1 + \log_2(1+j)}
\end{equation}

\noindent ensuring that ${LRE}/C_{LRE} \leq 1$. Note the ``floor'' function of the angular brackets above $\sum$ in Equation \ref{eq:clre}, which rounds the contents of the brackets down to the next (lowest) integer. 

The somewhat involved definition of $C_{LRE}$ is due to the fact that we wish the maximal possible error attainable (i.e., rankings that produce the largest
possibly credibility \emph{and} relevance errors) to correspond to a value of $1$ for ${LRE}/C_{LRE}$. Observe that $NLRE$ is $1$ if no errors of any kind occur (because, in that case,
LRE is $0$).

Our NLRE measure satisfies the desiderata presented in Section \ref{s:desiderata} as follows:
\begin{itemize}

\item \textbf{D1} holds if we interpret error size as the size of the rank differences;

\item \textbf{D2} holds due to the discount factor of $1/\log_2(1+i)$;

\item \textbf{D3} is satisfied in the sense that larger differences in \emph{credibility} or \emph{relevance} ranks mean larger error;

\item \textbf{D4}: The credibility error is scaled by the relevance error, if there is any (i.e., they are multiplied). If there is no relevance error, the credibility error is still strictly greater than zero;

\item \textbf{D5}: The measure is well-defined in all cases; 

\item \textbf{D6}: No scores occur explicitly, only rankings, so scaling makes no difference;

\item \textbf{D7} is satisfied because if all documents have equal relevance, the relevance error will be zero. The resulting score will measure only credibility error. And vice versa;

\item \textbf{D8} is satisfied through $\mu$ and $\nu$. 

\end{itemize}

We call NLRE a \textit{local} measure because it is affected by differences in credibility and relevance between documents at each rank position in the input ranking. We present next a \textit{global} evaluation metric that does not take such ``local'' effects at each rank into account (i.e., any differences in credibility and relevance between documents at rank $i$ in the input ranking do not affect the global metric; only the total difference of credibility and relevance of the entire input ranking affects the global metric).

\subsection{Normalised Global Rank Error (NGRE)}
%


We define the Global Rank Error (GRE) evaluation measure as $GRE = 0$ if $n = 1$, and otherwise:

{\small
\begin{equation}
GRE = \left(1 + \mu \sum_{i=1}^{n-1}  \frac{1}{\log_2 (1+i)} \epsilon^r \right) \left(1 + \nu \sum_{i=1}^{n-1}  \frac{1}{\log_2 (1+i)} \epsilon^c \right) - 1
\end{equation}
}


\noindent The notation is the same as for LRE. Similarly to LRE, we invert and normalise GRE, to produce its normalised version (NGRE) as follows:

\begin{equation}
\label{eq:ngre}
NGRE = 1 - \frac{GRE}{C_{GRE}}
\end{equation}
\noindent where $C_{GRE}$ is the normalisation constant, defined as:

{\footnotesize
\begin{equation} \label{eq:cgre}
C_{GRE} = \mu\nu \left(  \sum_{j=0}^{\lfloor \frac{n}{2} -1 \rfloor} \frac{n - 2j - 1}{1 + \log_2(1+j)} \right)^2 + (\mu + \nu) \sum_{j=0}^{\lfloor \frac{n}{2} -1 \rfloor} \frac{n - 2j - 1}{1 + \log_2(1+j)}
\end{equation}
}
\noindent $C_{GRE}$ is chosen to ensure that $GRE/C_{GRE}  \leq 1$ and that $GRE/C_{GRE} = 1$ is possible if{f} the ranking has the maximal possible errors compared to both the ideal relevance and ideal
credibility rankings. The square brackets above both $\sum$s in Equation \ref{eq:cgre} also use the floor function, exactly like in Equation \ref{eq:clre}.



As with NLRE, NGRE is $1$ if no errors of any kind occur. In spite of the differences in computation, NGRE satisfies 
all eight desiderata for the same reasons given for NLRE.


The main intuitive difference between NLRE and NGRE is that in NGRE the credibility errors and relevance errors are cumulated separately, and then multiplied at the end. Thus, there is no immediate connection between credibility and relevance errors at the same rank (\textit{locally}), hence we say that the metric is \emph{global}.

The advantage of such a global versus local measure is that, in the global case, it is more straightforward to perform mathematical manipulations to achieve, e.g., normalisation, and easier to intuitively grasp
what the measure means. The disadvantage is that local information is lost, and this may, in theory, lead to poorly performing measures. As the notion of ``error'' defined earlier is inherently a local phenomenon, the desiderata concerning errors are harder to satisfy formally for global measures.

\section{Type II: Document Score Measures}
\label{s:scores}
The two evaluation measures presented above (NLRE and NGRE) operate on the rank positions of documents. We now present three evaluation measures that operate, not on the rank positions of documents, but directly on document scores.

\subsection{Normalised Weighted Cumulative Score (NWCS)}
Given a ranking of documents that we wish to evaluate, let $Z^r(i)$ denote the relevance score with respect to some query of the document ranked at position $i$, and let $Z^c(i)$ denote the credibility score of the document ranked at position $i$. Then, we define the Weighted Cumulative Score (WCS) measure as:

\begin{equation}
\label{eq:wcs}
WCS = \sum_{i=1}^{n} \frac{1}{\log_2(1+i)} (\lambda Z^r(i) + (1-\lambda) Z^c(i))
\end{equation}

\noindent where $n$ is the total number of documents in the ranking list, and $\lambda$ is a real number in $[0,1]$ controlling the impact of relevance versus credibility in the computation. We normalise WCS by dividing it by the value obtained by an ``ideal'' ranking maximizing the value of WCS (this is inspired by the normalisation of the NDCG evaluation measure \cite{JarvelinK02}):

\begin{equation}
\label{eq:nwcs}
NWCS = \frac{WCS}{IWCS} 
\end{equation}

\noindent where IWCS is the \textit{ideal} WCS, i.e. the maximum WCS that can be obtained on an ideal ranking of the same documents.

NWCS uses a simple weighted combination of relevance or credibility \emph{scores} in the same manner as the metric $NGRE$, but is applicable directly to relevance or credibility \textit{scores} (instead of ranking positions). 

Our NWCS measure satisfies the following of the desiderata presented in Section \ref{s:desiderata}:
\begin{itemize}

\item \textbf{D1} is satisfied as both $Z^r$ and $Z^c$ occur linearly in WCS;

\item \textbf{D2} is satisfied
due to the logarithmic discounting for increasing rank positions;

\item \textbf{D3} is satisfied by design as both $Z^r$ and $Z^c$ occur directly in the formula for WCS;


\item \textbf{D5} is satisfied as the measure is well-defined in all cases; 

\item \textbf{D6} is satisfied due to normalization;

\item \textbf{D7} is satisfied because the contribution of the credibility scores (if all are equal) is just a constant
in each term (and vice versa if relevance scores are all equal);

\item \textbf{D8} is satisfied due to the presence of $\lambda$. 
\end{itemize}

Of all desiderata, only \textbf{D4} is not satisfied: there is no scaling of credibility errors based on relevance. 
Despite this, the advantage of NWCS is that it is interpretable in much the same way as NDCG.

The main idea of the next two measures is that any two separate measures of \emph{either} relevance \emph{or} credibility, but not both, can be combined into a single aggregating measure of relevance and credibility. 
We next present two such aggregating measures.

\subsection{Convex aggregating measure (CAM)}
We define the convex aggregating measure (CAM) of relevance and credibility as:

\begin{equation}
\label{eq:cam}
CAM = \lambda M^r + (1 - \lambda) M^c
\end{equation}

\noindent where $M^r$ and $M^c$ denote respectively any valid relevance and credibility evaluation measure, and $\lambda$ is a real number in $[0,1]$ controlling the impact of the individual relevance or credibility measure in the overall computation. CAM is normalized if both $M_r$ and $M^c$ are normalised.

Our CAM measure satisfies the following desiderata:
\begin{itemize}

\item \textbf{D1} is satisfied for the same reasons as NWCS;

\item \textbf{D2} is not satisfied in general;

\item \textbf{D3} is satisfied for the same reasons as NWCS;

\item \textbf{D4} is not satisfied in general;

\item \textbf{D5} is satisfied for the same reasons as NWCS; 

\item \textbf{D6} is not satisfied in general; it is satisfied if both $M^r$ and $M^c$ are scale-free;

\item \textbf{D7} is satisfied because the contribution of the credibility scores (if all are equal) is just a constant
in each term (and vice versa if relevance scores are all equal);

\item \textbf{D8} is satisfied for the same reasons as NWCS. 
\end{itemize}

With respect to \textbf{D2}, \textbf{D4}, and \textbf{D6} not being satisfied in general: The tradeoff in this case is that as CAM is just a convex combination of existing measures,
the scores are readily interpretable by anyone able to interpret $M^r$ and $M^c$ scores.

\subsection{Weighted harmonic mean aggregating measure ( (WHAM) or ``$F$-score for credibility and ranking'')}

We define the weighted harmonic mean aggregating measure (WHAM) as zero if either $M^r$ or $M^c$ is zero, and otherwise:

\begin{equation}
\label{eq:WHMA}
WHAM = \frac{1}{\lambda \frac{1}{M^r} + (1 - \lambda) \frac{1}{M^c}}
\end{equation}

\noindent where the notation is the same as for CAM in Equation \ref{eq:cam} above.
WHAM is the weighted harmonic mean of $M^r$ and $M^c$.  Observe that if $\lambda = 0.5$,
WHAM is simply the F-1 scores of $M^r$ and $M^c$. Note that WHAM is normalized if both $M_r$ and $M^c$ are normalised.

Similar definitions of metrics can be made that use other averages. For example, one can use the weighted arithmetic and geometric means instead of the harmonic mean.


Our WHAM measure satisfies the following desiderata:
\begin{itemize}

\item \textbf{D1} is satisfied for the same reasons as CAM;

\item \textbf{D2} is not satisfied in general;

\item \textbf{D3} is satisfied for the same reasons as CAM;

\item \textbf{D4} is not satisfied in general;

\item \textbf{D5} is satisfied for the same reasons as CAM; 

\item \textbf{D6} is not satisfied in general; it is satisfied if both $M^r$ and $M^c$ are scale-free;

\item \textbf{D7} is satisfied for the same reasons as CAM;

\item \textbf{D8} is satisfied for the same reasons as CAM. 
\end{itemize}


%

The primary advantage
of CAM and WHAM is that their definitions appeal to simple concepts already known to larger audiences (convex combinations and averages), and hence the measures are simple to state and interpret. The
consequent disadvantage is that this simplicity comes at the cost of not satisfying all desiderata.

We next present an empirical evaluation of all our measures.

\section{Evaluation}
\label{s:eval}
There are two main approaches for evaluating evaluation measures: 
\begin{description}
\item [Axiomatic] Define some general fundamental properties that a measure should adhere to, and then reflect on how many of these properties are satisfied by a new measure, and to what extent.
\item [Empirical] Present a drawback of existing standard and accepted measures, and illustrate how a new measure addresses this. Ideally, the new measure should generally correlate well with the existing measures, except for the problematic cases, where it should perform better \cite{KumarV10}. 
\end{description}

We have already conducted the axiomatic evaluation of our measures, having presented 8 fundamental properties they should adhere to (Desiderata in Section \ref{s:desiderata}), and having subsequently discussed each of our measures in relation to these fundamental properties in Sections \ref{s:ranks} - \ref{s:scores}. We now present the empirical evaluation. We first present our in-house dataset and experimental setup, and then our findings.

\subsection{Empirical Evaluation}

The goal is to determine how good our measures are at evaluating both relevance and credibility in ranked lists. We do this by comparing the scores of our measures to the scores of well-known relevance and separately credibility measures. This comparison is done on a small dataset that we create for the purposes of this work as follows\footnote{Our dataset is freely available here: \url{https://github.com/diku-irlab/A66}}.
We formulated 10 queries that we thought were likely to fetch results of various levels of credibility if submitted to a web search engine. These queries are shown in Table \ref{tab:queries}.
\begin{table}
  \caption{The 10 queries used in our experiments.}
  \label{tab:queries}
  \begin{tabular}{ r l}
    \toprule
    Query no.	& Query\\
     \midrule
   1 & Smoking not bad for health \\
   2 &  Princess Diana alive\\
   3 & Trump scientologist\\
   4 & UFO sightings\\
   5 & Loch Ness monster sightings\\
   6 & Vaccines bad for children \\
   7 & Time travel proof\\
   8 & Brexit illuminati\\
   9 & Climate change not dangerous\\
   10 & Digital tv surveillance\\
    \bottomrule
  \end{tabular}
\end{table}
We then recruited 10 assessors (1 MSc student, 5 PhD students, 3 postdocs, and 1 assistant professor, all within Computer Science, but none working on this project; 1 female, 9 males). Assessors were asked to submit each query to Google, and to assign separately a score of relevance and a score of credibility to each of the top 5 results. Assessors were instructed to use the same graded scale of relevance and credibility shown in the first column of Table \ref{tab:conversion}.

Assessors were asked to use their own understanding of relevance and credibility, and not to let relevance affect their assessment of credibility, or vice versa (relevance and credibility were to be treated as unrelated aspects). Assessors were instructed that, if they did not understand a query, or if they were unsure about the credibility of a result, they should open a separate browser and try to gather more information on the topic. Assessors received a nominal reward for their effort.

Even though assessors used the same queries, the top 5 results retrieved from Google per query were not always identical. Consequently, we compute our measures separately on each assessed ranking, and we report the arithmetic average. For NLRE and NGRE, we set $\mu=\nu=0.5$, meaning that relevance and credibility are weighted equally. Similarly, for NWCS, CAM, and WHAM, we set $\lambda=0.5$.

As no measures of \emph{both} relevance and credibility exist, we compare the score of our measures on the above dataset to the scores of:
\begin{itemize}
\item NDCG (for graded relevance), AP (for binary relevance);
\item F-1, G-measure (for binary credibility).
\end{itemize}
F-1 was introduced in Section \ref{s:rw} for relevance. We use it here to assess credibility, by defining its constituent precision and recall in terms of true/false positives/negatives (as is standard in classification evaluation). The G-measure is the geometric mean of precision and recall, which are defined as for F-1.

To render our graded assessments binary (for AP, F-1, G-measure), we use the conversion shown in Table \ref{tab:conversion}.

\begin{table}
  \caption{Conversion of graded assessments to binary. The same conversion is applied to both relevance and credibility assessments.}
  \label{tab:conversion}
  \begin{tabular}{ l l}
    \toprule
    Graded	& Binary\\
     \midrule
   1 (not at all)	& 0 (not at all)\\
   2 (marginally) & 0 (not at all) \\
   3 (medium) 	& 1 (completely)\\
   4 (completely) &1 (completely)\\
    \bottomrule
  \end{tabular}
\end{table}


\subsection{Findings}


Table \ref{tab:scores} displays the scores of all evaluation measures on our dataset. We see that relevance-only measures (NDCG, AP) give overall higher scores than credibility-only measures (F-1, G). It is not surprising to see such high NDCG and AP scores, considering that we assess only the top 5 ranks of Google. What is however interesting, is the comparatively lower scores of credibility (F-1 and G). This practically means that even the top ranks of a high-traffic web search engine like Google can be occupied by information that is not entirely credible (at least for this specially selected set of queries).

Looking at our measures of evaluation and credibility, we see that they range from roughly 0.6 to 0.9. This coincides with the range between the score of credibility-only measures and relevance-only measures. All of our measures are strongly and positively correlated to NDCG, AP, F-1, and G (from Spearman's $\rho$ = 0.79 for NDCG and F-1, up to $\rho$ = 0.97 for NDCG and NLRE). 

\begin{table}
  \caption{Our evaluation measures compared to NDCG, AP, F-1 and G. For NDCG we see our graded assessments. For the rest, we convert our graded assessments to binary as follows: 1 or 2 = not relevant/credible; 3 or 4 = relevant or credible. All measures are computed on the top 5 results returned for each query shown in Table \ref{tab:queries}. We report the average across all assessors.}
  \label{tab:scores}
  \begin{tabular}{ l r}
    \toprule
    \multicolumn{2}{c}{\bf{RELEVANCE}} \\
    NDCG &0.9329\\ 
    AP	&0.7842\\
     \midrule
   \multicolumn{2}{c}{\bf{CREDIBILITY}} \\
    F-1	&0.4786\\
    G	&0.5475\\
    \midrule
    \multicolumn{2}{c}{\bf{RELEVANCE and CREDIBILITY}} \\
    NLRE	&0.8262\\
    NGRE	&0.6919\\
    NWCS	&0.9413\\
    CAM$_{NDCG,F-1}$	&0.7058\\ 
    CAM$_{NDCG,G}$		&0.7402\\ 
    CAM$_{AP,F-1}$		&0.6311\\ 
    CAM$_{AP,G}$		&0.6659\\
    WHAM$_{NDCG,F-1}$	&0.6326\\
    WHAM$_{NDCG,G}$		&0.6900\\
    WHAM$_{AP,F-1}$		&0.6089\\
    WHAM$_{AP,G}$		&0.6448\\    
    \bottomrule
  \end{tabular}
\end{table}

\begin{table*}
  \caption{Examples of max/min relevance and credibility, from our experiments. Only one out of the 5 retrieved documents is shown per query. The urls of the retrieved results are reduced to their most content-bearing parts, for brevity.}
  \label{tab:perq}
  \begin{tabular}{r l c c c c c c c c c}
    \toprule
    \multicolumn{11}{c}{EXAMPLES OF HIGH RELEVANCE AND LOW CREDIBILITY}\\
    Query &Result (rank) &Relevance & Credibility &NDCG & AP & F-1 &G&NLRE &NGRE &NWCS\\
    \midrule
    2&{\footnotesize{www.surrealscoop.com$\ldots$princess-diana-found-alive}} (3)&4&1&.883&.679&.333&.387&.819&.585&.950\\
    3&{\footnotesize{tonyortega.org$\dots$scientology$\dots$where-does-trump-stand}} (1)&4&1&.938&1.00&.571&.631&.949&.797&.913\\
    4&{\footnotesize{www.ufosightingsdaily.com}} (1)&4&1&1.00&1.00&.333&.431&.808&.262&.941\\
    6&{\footnotesize{articles.mercola.com$\ldots$vaccines-adverse-reaction}} (4)&4&1&.938&.950&.571&.500&.872&.534&.927\\
    8&{\footnotesize{www.henrymakow.com$\ldots$brexit-what-is-the-globalist-game}} (1)&4&1&.884&.679&.000&.000&.889&.666&.985\\
    10&{\footnotesize{educate-yourself.org$\ldots$HDtvcovertsurveillanceagenda}} (3)&4&1&.979&1.00&.000&.000&.926&.885&.997\\

    \midrule
     \multicolumn{11}{c}{EXAMPLES OF HIGH CREDIBILITY AND LOW RELEVANCE}\\
    \midrule
     10&{\footnotesize{cctvcamerapros.com/Connect-CCTV-Camera-to-TV-s}} (2)&1&4&.780&.533&.571&.715&.863&.710&.931\\
     10&{\footnotesize{ieeexplore.ieee.org/document/891879}} (5)&1&4&.780&.533&.571&.715&.899&.605&.874\\
    \bottomrule
  \end{tabular}
\end{table*}

Table \ref{tab:perq} shows examples of high divergence between the relevance and credibility of the retrieved documents, for three of our measures (the scores of our remaining metrics can be easily deduced from the respective relevance-only and credibility-only scores, as our omitted measures -- CAM and WHAM -- aggregate the existing relevance-only and credibility-only metrics shown in Table \ref{tab:perq}). Note that, whereas we found several examples of max relevance and min credibility in our data, there were (understandably) significantly fewer examples of max credibility and min relevance (this distribution is reflected in Table \ref{tab:perq}). We see that NWCS gives higher scores for queries 2 and 4-10 than NLRE and NGRE. For the first five examples (of max relevance and min credibility), this is likely because NWCS does not satisfy \textbf{D4}, namely that credibility errors should be penalised more on high relevance versus low relevance documents. We also see that NGRE gives consistently lower scores than NLRE and NWCS. This is due to its \textit{global} aspect discussed earlier: NGRE accumulates credibility and relevance errors separately and then multiplies them at the end, meaning that local errors in each rank do not impact as much the final score (unlike NLRE and NWCS, which are both \textit{local} in that sense, the first using document ranks, the second using document scores). 

\section{Conclusions}

The credibility of search results is important in many retrieval tasks, and should be, we reason, integrated into IR evaluation measures
that are, as of now, targetting mostly relevance. We have presented several measures and types of measures that can be used to gauge the effectiveness of a ranking, taking into account both credibility and relevance. The measures are both axiomatically and empirically sound, the latter illustrated in a small user study.

There are at least two natural extensions of our approach: First, the combination of rankings based on different criteria goes beyond the combination of relevance and credibility, and several such combinations are used in practice based on different criteria (e.g., combinations of relevance and upvotes on social media sites); we believe that much of our work can be encompassed in more general approaches, suitably axiomatised, that do not necessarily have to satisfy the same desiderata as those of this paper (e.g., do not have to scale credibility error by relevance errors as in our \textbf{D4}). Second,
while we have chosen to devise measures that are both theoretically principled and conceptually simple using simple criteria (satisfaction of desiderata, local versus global, amenable to principled interpretation), there are many more measures that can be defined within the same limits. For example, our Type II measures are primarily built on simple combinations of scores or pre-existing measures that can easily be understood by the community, but at the price that some desiderata are hard or impossible to satisfy; however, there is no theoretical reason why one could not create Type II measures that incorporate some of the ideas from Type I metrics. We intend to investigate these two extensions in the future, and invite the community to do so as well.

Lastly, while the notion of credibility, in particular in news media, is subject to intense public discussion, very few empirical studies exist that contain user preferences, credibility rankings, or information needs related to credibility. The small study included in this paper, while informative, is a very small step in this direction. We believe that future substantial discussion of \emph{practically relevant} research involving credibility in information retrieval would greatly benefit from having access to larger-scale empirical user studies.

\bibliographystyle{ACM-Reference-Format}
\bibliography{ictircre} 


\begin{thebibliography}{00}


\ifx \showCODEN    \undefined \def \showCODEN     #1{\unskip}     \fi
\ifx \showDOI      \undefined \def \showDOI       #1{{\tt DOI:}\penalty0{#1}\ }
  \fi
\ifx \showISBNx    \undefined \def \showISBNx     #1{\unskip}     \fi
\ifx \showISBNxiii \undefined \def \showISBNxiii  #1{\unskip}     \fi
\ifx \showISSN     \undefined \def \showISSN      #1{\unskip}     \fi
\ifx \showLCCN     \undefined \def \showLCCN      #1{\unskip}     \fi
\ifx \shownote     \undefined \def \shownote      #1{#1}          \fi
\ifx \showarticletitle \undefined \def \showarticletitle #1{#1}   \fi
\ifx \showURL      \undefined \def \showURL       {\relax}        \fi
\providecommand\bibfield[2]{#2}
\providecommand\bibinfo[2]{#2}
\providecommand\natexlab[1]{#1}
\providecommand\showeprint[2][]{arXiv:#2}

\bibitem[\protect\citeauthoryear{Balakrishnan and Kambhampati}{Balakrishnan and
  Kambhampati}{2011}]%
        {BalakrishnanK11}
\bibfield{author}{\bibinfo{person}{Raju Balakrishnan} {and}
  \bibinfo{person}{Subbarao Kambhampati}.} \bibinfo{year}{2011}\natexlab{}.
\newblock \showarticletitle{SourceRank: relevance and trust assessment for deep
  web sources based on inter-source agreement}. In \bibinfo{booktitle}{{\em
  Proceedings of the 20th International Conference on World Wide Web, {WWW}
  2011, Hyderabad, India, March 28 - April 1, 2011}},
  \bibfield{editor}{\bibinfo{person}{Sadagopan Srinivasan},
  \bibinfo{person}{Krithi Ramamritham}, \bibinfo{person}{Arun Kumar},
  \bibinfo{person}{M.~P. Ravindra}, \bibinfo{person}{Elisa Bertino}, {and}
  \bibinfo{person}{Ravi Kumar}} (Eds.). \bibinfo{publisher}{{ACM}},
  \bibinfo{pages}{227--236}.
\newblock
\showISBNx{978-1-4503-0632-4}
\showDOI{%
\url{https://doi.org/10.1145/1963405.1963440}}


\bibitem[\protect\citeauthoryear{Black, Scarfone, and Souppaya}{Black
  et~al\mbox{.}}{2008}]%
        {BlackSS08}
\bibfield{editor}{\bibinfo{person}{Paul~E. Black}, \bibinfo{person}{Karen~A.
  Scarfone}, {and} \bibinfo{person}{Murugiah~P. Souppaya}} (Eds.).
  \bibinfo{year}{2008}\natexlab{}.
\newblock \bibinfo{booktitle}{{\em Cyber Security Metrics and Measures}}.
  \bibinfo{publisher}{{Wiley Handbook of Science and Technology for Homeland
  Security}}.
\newblock


\bibitem[\protect\citeauthoryear{Brost, Cox, Seldin, and Lioma}{Brost
  et~al\mbox{.}}{2016a}]%
        {BrostCSL16a}
\bibfield{author}{\bibinfo{person}{Brian Brost}, \bibinfo{person}{Ingemar~J.
  Cox}, \bibinfo{person}{Yevgeny Seldin}, {and} \bibinfo{person}{Christina
  Lioma}.} \bibinfo{year}{2016}\natexlab{a}.
\newblock \showarticletitle{An Improved Multileaving Algorithm for Online
  Ranker Evaluation}. In \bibinfo{booktitle}{{\em Proceedings of the 39th
  International {ACM} {SIGIR} conference on Research and Development in
  Information Retrieval, {SIGIR} 2016, Pisa, Italy, July 17-21, 2016}},
  \bibfield{editor}{\bibinfo{person}{Raffaele Perego},
  \bibinfo{person}{Fabrizio Sebastiani}, \bibinfo{person}{Javed~A. Aslam},
  \bibinfo{person}{Ian Ruthven}, {and} \bibinfo{person}{Justin Zobel}} (Eds.).
  \bibinfo{publisher}{{ACM}}, \bibinfo{pages}{745--748}.
\newblock
\showISBNx{978-1-4503-4069-4}
\showDOI{%
\url{https://doi.org/10.1145/2911451.2914706}}


\bibitem[\protect\citeauthoryear{Brost, Seldin, Cox, and Lioma}{Brost
  et~al\mbox{.}}{2016b}]%
        {BrostSCL16}
\bibfield{author}{\bibinfo{person}{Brian Brost}, \bibinfo{person}{Yevgeny
  Seldin}, \bibinfo{person}{Ingemar~J. Cox}, {and} \bibinfo{person}{Christina
  Lioma}.} \bibinfo{year}{2016}\natexlab{b}.
\newblock \showarticletitle{Multi-Dueling Bandits and Their Application to
  Online Ranker Evaluation}. In \bibinfo{booktitle}{{\em Proceedings of the
  25th {ACM} International on Conference on Information and Knowledge
  Management, {CIKM} 2016, Indianapolis, IN, USA, October 24-28, 2016}},
  \bibfield{editor}{\bibinfo{person}{Snehasis Mukhopadhyay},
  \bibinfo{person}{ChengXiang Zhai}, \bibinfo{person}{Elisa Bertino},
  \bibinfo{person}{Fabio Crestani}, \bibinfo{person}{Javed Mostafa},
  \bibinfo{person}{Jie Tang}, \bibinfo{person}{Luo Si},
  \bibinfo{person}{Xiaofang Zhou}, \bibinfo{person}{Yi~Chang},
  \bibinfo{person}{Yunyao Li}, {and} \bibinfo{person}{Parikshit Sondhi}}
  (Eds.). \bibinfo{publisher}{{ACM}}, \bibinfo{pages}{2161--2166}.
\newblock


\bibitem[\protect\citeauthoryear{Buckley and Voorhees}{Buckley and
  Voorhees}{2004}]%
        {BuckleyV04}
\bibfield{author}{\bibinfo{person}{Chris Buckley} {and}
  \bibinfo{person}{Ellen~M. Voorhees}.} \bibinfo{year}{2004}\natexlab{}.
\newblock \showarticletitle{Retrieval evaluation with incomplete information}.
  In \bibinfo{booktitle}{{\em {SIGIR} 2004: Proceedings of the 27th Annual
  International {ACM} {SIGIR} Conference on Research and Development in
  Information Retrieval, Sheffield, UK, July 25-29, 2004}},
  \bibfield{editor}{\bibinfo{person}{Mark Sanderson}, \bibinfo{person}{Kalervo
  J{\"{a}}rvelin}, \bibinfo{person}{James Allan}, {and} \bibinfo{person}{Peter
  Bruza}} (Eds.). \bibinfo{publisher}{{ACM}}, \bibinfo{pages}{25--32}.
\newblock
\showISBNx{1-58113-881-4}
\showDOI{%
\url{https://doi.org/10.1145/1008992.1009000}}


\bibitem[\protect\citeauthoryear{Chapelle, Metlzer, Zhang, and
  Grinspan}{Chapelle et~al\mbox{.}}{2009}]%
        {ChapelleMZG09}
\bibfield{author}{\bibinfo{person}{Olivier Chapelle}, \bibinfo{person}{Donald
  Metlzer}, \bibinfo{person}{Ya Zhang}, {and} \bibinfo{person}{Pierre
  Grinspan}.} \bibinfo{year}{2009}\natexlab{}.
\newblock \showarticletitle{Expected reciprocal rank for graded relevance}. In
  \bibinfo{booktitle}{{\em Proceedings of the 18th {ACM} Conference on
  Information and Knowledge Management, {CIKM} 2009, Hong Kong, China, November
  2-6, 2009}}, \bibfield{editor}{\bibinfo{person}{David~Wai{-}Lok Cheung},
  \bibinfo{person}{Il{-}Yeol Song}, \bibinfo{person}{Wesley~W. Chu},
  \bibinfo{person}{Xiaohua Hu}, {and} \bibinfo{person}{Jimmy~J. Lin}} (Eds.).
  \bibinfo{publisher}{{ACM}}, \bibinfo{pages}{621--630}.
\newblock
\showISBNx{978-1-60558-512-3}
\showDOI{%
\url{https://doi.org/10.1145/1645953.1646033}}


\bibitem[\protect\citeauthoryear{Chuklin, Serdyukov, and de~Rijke}{Chuklin
  et~al\mbox{.}}{2013}]%
        {ChuklinSR13}
\bibfield{author}{\bibinfo{person}{Aleksandr Chuklin}, \bibinfo{person}{Pavel
  Serdyukov}, {and} \bibinfo{person}{Maarten de Rijke}.}
  \bibinfo{year}{2013}\natexlab{}.
\newblock \showarticletitle{Click model-based information retrieval metrics}.
  In \bibinfo{booktitle}{{\em The 36th International {ACM} {SIGIR} conference
  on research and development in Information Retrieval, {SIGIR} '13, Dublin,
  Ireland - July 28 - August 01, 2013}},
  \bibfield{editor}{\bibinfo{person}{Gareth J.~F. Jones},
  \bibinfo{person}{Paraic Sheridan}, \bibinfo{person}{Diane Kelly},
  \bibinfo{person}{Maarten de~Rijke}, {and} \bibinfo{person}{Tetsuya Sakai}}
  (Eds.). \bibinfo{publisher}{{ACM}}, \bibinfo{pages}{493--502}.
\newblock
\showISBNx{978-1-4503-2034-4}
\showDOI{%
\url{https://doi.org/10.1145/2484028.2484071}}


\bibitem[\protect\citeauthoryear{Ennals, Byler, Agosta, and Rosario}{Ennals
  et~al\mbox{.}}{2010}]%
        {EnnalsBAR10}
\bibfield{author}{\bibinfo{person}{Rob Ennals}, \bibinfo{person}{Dan Byler},
  \bibinfo{person}{John~Mark Agosta}, {and} \bibinfo{person}{Barbara Rosario}.}
  \bibinfo{year}{2010}\natexlab{}.
\newblock \showarticletitle{What is disputed on the web?}. In
  \bibinfo{booktitle}{{\em Proceedings of the 4th {ACM} Workshop on Information
  Credibility on the Web, {WICOW} 2010, Raleigh, North Carolina, USA, April 27,
  2010}}, \bibfield{editor}{\bibinfo{person}{Katsumi Tanaka},
  \bibinfo{person}{Xiaofang Zhou}, \bibinfo{person}{Min Zhang}, {and}
  \bibinfo{person}{Adam Jatowt}} (Eds.). \bibinfo{publisher}{{ACM}},
  \bibinfo{pages}{67--74}.
\newblock
\showISBNx{978-1-60558-940-4}
\showDOI{%
\url{https://doi.org/10.1145/1772938.1772952}}


\bibitem[\protect\citeauthoryear{Hofmann, Li, and Radlinski}{Hofmann
  et~al\mbox{.}}{2016}]%
        {HofmannLR16}
\bibfield{author}{\bibinfo{person}{Katja Hofmann}, \bibinfo{person}{Lihong Li},
  {and} \bibinfo{person}{Filip Radlinski}.} \bibinfo{year}{2016}\natexlab{}.
\newblock \showarticletitle{Online Evaluation for Information Retrieval}.
\newblock \bibinfo{journal}{{\em Foundations and Trends in Information
  Retrieval\/}} \bibinfo{volume}{10}, \bibinfo{number}{1}
  (\bibinfo{year}{2016}), \bibinfo{pages}{1--117}.
\newblock


\bibitem[\protect\citeauthoryear{Horn, Zhila, Gelbukh, Kern, and Lex}{Horn
  et~al\mbox{.}}{2013}]%
        {HornZGKL13}
\bibfield{author}{\bibinfo{person}{Christopher Horn}, \bibinfo{person}{Alisa
  Zhila}, \bibinfo{person}{Alexander~F. Gelbukh}, \bibinfo{person}{Roman Kern},
  {and} \bibinfo{person}{Elisabeth Lex}.} \bibinfo{year}{2013}\natexlab{}.
\newblock \showarticletitle{Using Factual Density to Measure Informativeness of
  Web Documents}. In \bibinfo{booktitle}{{\em Proceedings of the 19th Nordic
  Conference of Computational Linguistics, {NODALIDA} 2013, May 22-24, 2013,
  Oslo University, Norway}} {\em (\bibinfo{series}{Link{\"{o}}ping Electronic
  Conference Proceedings})}, \bibfield{editor}{\bibinfo{person}{Stephan Oepen},
  \bibinfo{person}{Kristin Hagen}, {and} \bibinfo{person}{Janne~Bondi
  Johannessen}} (Eds.), Vol.~\bibinfo{volume}{85}.
  \bibinfo{publisher}{Link{\"{o}}ping University Electronic Press},
  \bibinfo{pages}{227--238}.
\newblock
\showURL{%
\url{http://www.ep.liu.se/ecp_article/index.en.aspx?issue=085};
\url{article=021}}


\bibitem[\protect\citeauthoryear{Huang, Olteanu, and Aberer}{Huang
  et~al\mbox{.}}{2013}]%
        {HuangOA13}
\bibfield{author}{\bibinfo{person}{Zhicong Huang}, \bibinfo{person}{Alexandra
  Olteanu}, {and} \bibinfo{person}{Karl Aberer}.}
  \bibinfo{year}{2013}\natexlab{}.
\newblock \showarticletitle{CredibleWeb: a platform for web credibility
  evaluation}. In \bibinfo{booktitle}{{\em 2013 {ACM} {SIGCHI} Conference on
  Human Factors in Computing Systems, {CHI} '13, Paris, France, April 27 - May
  2, 2013, Extended Abstracts}}, \bibfield{editor}{\bibinfo{person}{Wendy~E.
  Mackay}, \bibinfo{person}{Stephen~A. Brewster}, {and}
  \bibinfo{person}{Susanne B{\o}dker}} (Eds.). \bibinfo{publisher}{{ACM}},
  \bibinfo{pages}{1887--1892}.
\newblock


\bibitem[\protect\citeauthoryear{J{\"{a}}rvelin and
  Kek{\"{a}}l{\"{a}}inen}{J{\"{a}}rvelin and Kek{\"{a}}l{\"{a}}inen}{2002}]%
        {JarvelinK02}
\bibfield{author}{\bibinfo{person}{Kalervo J{\"{a}}rvelin} {and}
  \bibinfo{person}{Jaana Kek{\"{a}}l{\"{a}}inen}.}
  \bibinfo{year}{2002}\natexlab{}.
\newblock \showarticletitle{Cumulated gain-based evaluation of {IR}
  techniques}.
\newblock \bibinfo{journal}{{\em {ACM} Trans. Inf. Syst.\/}}
  \bibinfo{volume}{20}, \bibinfo{number}{4} (\bibinfo{year}{2002}),
  \bibinfo{pages}{422--446}.
\newblock
\showDOI{%
\url{https://doi.org/10.1145/582415.582418}}


\bibitem[\protect\citeauthoryear{Kumar and Vassilvitskii}{Kumar and
  Vassilvitskii}{2010}]%
        {KumarV10}
\bibfield{author}{\bibinfo{person}{Ravi Kumar} {and} \bibinfo{person}{Sergei
  Vassilvitskii}.} \bibinfo{year}{2010}\natexlab{}.
\newblock \showarticletitle{Generalized distances between rankings}. In
  \bibinfo{booktitle}{{\em Proceedings of the 19th International Conference on
  World Wide Web, {WWW} 2010, Raleigh, North Carolina, USA, April 26-30,
  2010}}, \bibfield{editor}{\bibinfo{person}{Michael Rappa},
  \bibinfo{person}{Paul Jones}, \bibinfo{person}{Juliana Freire}, {and}
  \bibinfo{person}{Soumen Chakrabarti}} (Eds.). \bibinfo{publisher}{{ACM}},
  \bibinfo{pages}{571--580}.
\newblock
\showISBNx{978-1-60558-799-8}
\showDOI{%
\url{https://doi.org/10.1145/1772690.1772749}}


\bibitem[\protect\citeauthoryear{Lex, Khan, Bischof, and Granitzer}{Lex
  et~al\mbox{.}}{2014}]%
        {LexKBG14}
\bibfield{author}{\bibinfo{person}{Elisabeth Lex}, \bibinfo{person}{Inayat
  Khan}, \bibinfo{person}{Horst Bischof}, {and} \bibinfo{person}{Michael
  Granitzer}.} \bibinfo{year}{2014}\natexlab{}.
\newblock \showarticletitle{Assessing the Quality of Web Content}.
\newblock \bibinfo{journal}{{\em CoRR\/}}  \bibinfo{volume}{abs/1406.3188}
  (\bibinfo{year}{2014}).
\newblock
\showURL{%
\url{http://arxiv.org/abs/1406.3188}}


\bibitem[\protect\citeauthoryear{Lioma, Larsen, Lu, and Huang}{Lioma
  et~al\mbox{.}}{2016}]%
        {LiomaLLH16}
\bibfield{author}{\bibinfo{person}{Christina Lioma}, \bibinfo{person}{Birger
  Larsen}, \bibinfo{person}{Wei Lu}, {and} \bibinfo{person}{Yong Huang}.}
  \bibinfo{year}{2016}\natexlab{}.
\newblock \showarticletitle{A study of factuality, objectivity and relevance:
  three desiderata in large-scale information retrieval?}. In
  \bibinfo{booktitle}{{\em Proceedings of the 3rd {IEEE/ACM} International
  Conference on Big Data Computing, Applications and Technologies, {BDCAT}
  2016, Shanghai, China, December 6-9, 2016}},
  \bibfield{editor}{\bibinfo{person}{Ashiq Anjum} {and}
  \bibinfo{person}{Xinghui Zhao}} (Eds.). \bibinfo{publisher}{{ACM}},
  \bibinfo{pages}{107--117}.
\newblock
\showISBNx{978-1-4503-4617-7}
\showDOI{%
\url{https://doi.org/10.1145/3006299.3006315}}


\bibitem[\protect\citeauthoryear{Mizzaro}{Mizzaro}{2008}]%
        {Mizzaro08}
\bibfield{author}{\bibinfo{person}{Stefano Mizzaro}.}
  \bibinfo{year}{2008}\natexlab{}.
\newblock \showarticletitle{The Good, the Bad, the Difficult, and the Easy:
  Something Wrong with Information Retrieval Evaluation?}. In
  \bibinfo{booktitle}{{\em Advances in Information Retrieval , 30th European
  Conference on {IR} Research, {ECIR} 2008, Glasgow, UK, March 30-April 3,
  2008. Proceedings}} {\em (\bibinfo{series}{Lecture Notes in Computer
  Science})}, \bibfield{editor}{\bibinfo{person}{Craig Macdonald},
  \bibinfo{person}{Iadh Ounis}, \bibinfo{person}{Vassilis Plachouras},
  \bibinfo{person}{Ian Ruthven}, {and} \bibinfo{person}{Ryen~W. White}} (Eds.),
  Vol.~\bibinfo{volume}{4956}. \bibinfo{publisher}{Springer},
  \bibinfo{pages}{642--646}.
\newblock
\showISBNx{978-3-540-78645-0}
\showDOI{%
\url{https://doi.org/10.1007/978-3-540-78646-7_71}}


\bibitem[\protect\citeauthoryear{Morris, Counts, Roseway, Hoff, and
  Schwarz}{Morris et~al\mbox{.}}{2012}]%
        {MorrisCRHS12}
\bibfield{author}{\bibinfo{person}{Meredith~Ringel Morris},
  \bibinfo{person}{Scott Counts}, \bibinfo{person}{Asta Roseway},
  \bibinfo{person}{Aaron Hoff}, {and} \bibinfo{person}{Julia Schwarz}.}
  \bibinfo{year}{2012}\natexlab{}.
\newblock \showarticletitle{Tweeting is believing?: understanding microblog
  credibility perceptions}. In \bibinfo{booktitle}{{\em {CSCW} '12 Computer
  Supported Cooperative Work, Seattle, WA, USA, February 11-15, 2012}},
  \bibfield{editor}{\bibinfo{person}{Steven~E. Poltrock},
  \bibinfo{person}{Carla Simone}, \bibinfo{person}{Jonathan Grudin},
  \bibinfo{person}{Gloria Mark}, {and} \bibinfo{person}{John Riedl}} (Eds.).
  \bibinfo{publisher}{{ACM}}, \bibinfo{pages}{441--450}.
\newblock
\showISBNx{978-1-4503-1086-4}
\showDOI{%
\url{https://doi.org/10.1145/2145204.2145274}}


\bibitem[\protect\citeauthoryear{Park, Kang, Chung, and Song}{Park
  et~al\mbox{.}}{2009}]%
        {ParkKCS09}
\bibfield{author}{\bibinfo{person}{Souneil Park}, \bibinfo{person}{Seungwoo
  Kang}, \bibinfo{person}{Sangyoung Chung}, {and} \bibinfo{person}{Junehwa
  Song}.} \bibinfo{year}{2009}\natexlab{}.
\newblock \showarticletitle{NewsCube: delivering multiple aspects of news to
  mitigate media bias}. In \bibinfo{booktitle}{{\em Proceedings of the 27th
  International Conference on Human Factors in Computing Systems, {CHI} 2009,
  Boston, MA, USA, April 4-9, 2009}}, \bibfield{editor}{\bibinfo{person}{Dan
  R.~Olsen Jr.}, \bibinfo{person}{Richard~B. Arthur}, \bibinfo{person}{Ken
  Hinckley}, \bibinfo{person}{Meredith~Ringel Morris},
  \bibinfo{person}{Scott~E. Hudson}, {and} \bibinfo{person}{Saul Greenberg}}
  (Eds.). \bibinfo{publisher}{{ACM}}, \bibinfo{pages}{443--452}.
\newblock
\showISBNx{978-1-60558-246-7}


\bibitem[\protect\citeauthoryear{Schuth, Hofmann, and Radlinski}{Schuth
  et~al\mbox{.}}{2015}]%
        {SchuthHR15}
\bibfield{author}{\bibinfo{person}{Anne Schuth}, \bibinfo{person}{Katja
  Hofmann}, {and} \bibinfo{person}{Filip Radlinski}.}
  \bibinfo{year}{2015}\natexlab{}.
\newblock \showarticletitle{Predicting Search Satisfaction Metrics with
  Interleaved Comparisons}. In \bibinfo{booktitle}{{\em Proceedings of the 38th
  International {ACM} {SIGIR} Conference on Research and Development in
  Information Retrieval, Santiago, Chile, August 9-13, 2015}},
  \bibfield{editor}{\bibinfo{person}{Ricardo~A. Baeza{-}Yates},
  \bibinfo{person}{Mounia Lalmas}, \bibinfo{person}{Alistair Moffat}, {and}
  \bibinfo{person}{Berthier~A. Ribeiro{-}Neto}} (Eds.).
  \bibinfo{publisher}{{ACM}}, \bibinfo{pages}{463--472}.
\newblock


\bibitem[\protect\citeauthoryear{Schuth, Oosterhuis, Whiteson, and
  de~Rijke}{Schuth et~al\mbox{.}}{2016}]%
        {SchuthOWR16}
\bibfield{author}{\bibinfo{person}{Anne Schuth}, \bibinfo{person}{Harrie
  Oosterhuis}, \bibinfo{person}{Shimon Whiteson}, {and}
  \bibinfo{person}{Maarten de Rijke}.} \bibinfo{year}{2016}\natexlab{}.
\newblock \showarticletitle{Multileave Gradient Descent for Fast Online
  Learning to Rank}. In \bibinfo{booktitle}{{\em Proceedings of the Ninth {ACM}
  International Conference on Web Search and Data Mining, San Francisco, CA,
  USA, February 22-25, 2016}}. \bibinfo{pages}{457--466}.
\newblock


\bibitem[\protect\citeauthoryear{Schwarz and Morris}{Schwarz and
  Morris}{2011}]%
        {SchwarzM11}
\bibfield{author}{\bibinfo{person}{Julia Schwarz} {and}
  \bibinfo{person}{Meredith~Ringel Morris}.} \bibinfo{year}{2011}\natexlab{}.
\newblock \showarticletitle{Augmenting web pages and search results to support
  credibility assessment}. In \bibinfo{booktitle}{{\em Proceedings of the
  International Conference on Human Factors in Computing Systems, {CHI} 2011,
  Vancouver, BC, Canada, May 7-12, 2011}},
  \bibfield{editor}{\bibinfo{person}{Desney~S. Tan}, \bibinfo{person}{Saleema
  Amershi}, \bibinfo{person}{Bo~Begole}, \bibinfo{person}{Wendy~A. Kellogg},
  {and} \bibinfo{person}{Manas Tungare}} (Eds.). \bibinfo{publisher}{{ACM}},
  \bibinfo{pages}{1245--1254}.
\newblock
\showISBNx{978-1-4503-0228-9}


\bibitem[\protect\citeauthoryear{van Rijsbergen}{van Rijsbergen}{1974}]%
        {keith74}
\bibfield{author}{\bibinfo{person}{C.~J.~Keith van Rijsbergen}.}
  \bibinfo{year}{1974}\natexlab{}.
\newblock \showarticletitle{Foundation of evaluation}.
\newblock \bibinfo{journal}{{\em Journal of Documentation\/}}
  \bibinfo{volume}{30}, \bibinfo{number}{4} (\bibinfo{year}{1974}),
  \bibinfo{pages}{365--373}.
\newblock


\bibitem[\protect\citeauthoryear{Wiebe and Riloff}{Wiebe and Riloff}{2011}]%
        {WiebeR11}
\bibfield{author}{\bibinfo{person}{Janyce Wiebe} {and} \bibinfo{person}{Ellen
  Riloff}.} \bibinfo{year}{2011}\natexlab{}.
\newblock \showarticletitle{Finding Mutual Benefit between Subjectivity
  Analysis and Information Extraction}.
\newblock \bibinfo{journal}{{\em {IEEE} Trans. Affective Computing\/}}
  \bibinfo{volume}{2}, \bibinfo{number}{4} (\bibinfo{year}{2011}),
  \bibinfo{pages}{175--191}.
\newblock
\showDOI{%
\url{https://doi.org/10.1109/T-AFFC.2011.19}}


\bibitem[\protect\citeauthoryear{Yilmaz, Shokouhi, Craswell, and
  Robertson}{Yilmaz et~al\mbox{.}}{2010}]%
        {YilmazSCR10}
\bibfield{author}{\bibinfo{person}{Emine Yilmaz}, \bibinfo{person}{Milad
  Shokouhi}, \bibinfo{person}{Nick Craswell}, {and} \bibinfo{person}{Stephen
  Robertson}.} \bibinfo{year}{2010}\natexlab{}.
\newblock \showarticletitle{Expected browsing utility for web search
  evaluation}. In \bibinfo{booktitle}{{\em Proceedings of the 19th {ACM}
  Conference on Information and Knowledge Management, {CIKM} 2010, Toronto,
  Ontario, Canada, October 26-30, 2010}},
  \bibfield{editor}{\bibinfo{person}{Jimmy Huang}, \bibinfo{person}{Nick
  Koudas}, \bibinfo{person}{Gareth J.~F. Jones}, \bibinfo{person}{Xindong Wu},
  \bibinfo{person}{Kevyn Collins{-}Thompson}, {and} \bibinfo{person}{Aijun An}}
  (Eds.). \bibinfo{publisher}{{ACM}}, \bibinfo{pages}{1561--1564}.
\newblock
\showISBNx{978-1-4503-0099-5}
\showDOI{%
\url{https://doi.org/10.1145/1871437.1871672}}


\end{thebibliography}

\end{document}